# Continuous first order orbital order-disorder transition in $Nd_{1-x}Ca_xMnO_3$

C. V. Colin[1], A.J.C. Buurma[1], M. v. Zimmermann[2], T.T.M. Palstra[1]

[1] *Zernike Institue for Advanced Materials, University of Groningen, The Netherlands*
[2] *Hamburger Synchrotronstrahlungslabor HASYLAB at Deutsches Elektronen-Synchrotron DESY, 22603 Hamburg, Germany*

**Abstract**

The nature of the cooperative Jahn-Teller (JT) transition accompanied by orbital order-disorder in $Nd_{1-x}Ca_xMnO_3$ has been studied by high temperature (300 - 1200K) synchrotron and laboratory X-ray powder diffraction in the low doping region ($0 \leq x \leq 0.1$). For very low doping a large temperature range of phase coexistence associated with a first-order transition has been observed, resembling a martensitic transformation. The transition appears continuous because of a gradual evolution of the volume fractions of the phases over a broad temperature interval. The first-order nature of the transition and the phase coexistence is suppressed with increasing of doping.

## Introduction

Orbital physics is a key concept of the science and technology of transition metal oxides (see for example, Tokura *et al* 2000 and references therein). The orbital degree of freedom in combination with charge and spin determine the physical properties in hole-doped manganites. The lifting of the orbital degeneracy sets an energy scale intermediate to that of the crystal structure and the spin structure. The antisymmetrization of the total electronic wavefunction clearly implies that a strong relationship exists between the orbital and spin structures. The orbital state plays a crucial role in electronic properties. For instance, investigations of Ca doping in $LaMnO_3$ showed that the transition from an insulator to a ferromagnetic metal is not driven by the spin-exchange interactions but instead by the suppression of long-range cooperative Jahn-Teller, JT, distortion or orbital ordering by doping with a critical concentration near 21% (Van Aken *et al* 2003).

Order-disorder phase transitions, in particular those involving spins, are generally second order. However, the orbital ordering transition is associated with a large structural distortion and can be first order. There exists a general consensus about the 1st order nature of the orbital ordering transition for the parent compound $LaMnO_3$. However, there is no consensus on the order parameter, and the transition has mostly been probed using indirect parameters such as asymmetry in the distortion of Mn-O bond-lengths, or molar volume. There are presently no convenient probes to measure the orbital order parameter directly. This situation is further complicated for the manganite perovskites, where the orbital ordered state is isostructural in the spacegroup Pbnm, and occurs at elevated temperatures, $T_{OO}$>700'C. The large Debye-Waller factor and low electron density of oxygen make accurate structure determinations at these high temperatures time consuming for synchrotron/neutron diffraction measurements and often impossible for laboratory diffraction.

$LaMnO_3$ has been studied structurally in powder diffraction experiments by Rodriguez-Carvajal *et al* (1998), Prado *et al* (1999a) and Chaterji *et al* (2003). They observed a phase transition in $LaMnO_3$ from an orbital ordered orthorhombic O' phase at low temperature to an orbital disordered orthorhombic O phase at elevated temperature where the crystallographic symmetry is retained. Although the O' and O phases have the same crystallographic space group, they differ in molar volume. Furthermore, the low temperature O' phase shows a staggered $e_g$ orbital pattern in the basal plane which results in alternating Mn-O bond lengths associated with the Jahn-Teller effect. Static Jahn-Teller distortions become suppressed at high temperature where dynamic orbital ordering can still survive, as evidenced by local atomic structure probe techniques (Sanchez *et al* 2003, Qiu *et al* 2005). Rodriguez-Carvajal *et al* (1998) found an endothermic peak using DTA at the orbital ordering transition. The presence of latent heat for $LaMnO_3$ establishes the first order nature of the orbital order-disorder transition.

The orbital order-disorder transition in $LaMnO_3$ was studied by Zhou and Goodenough (2003) using electric and thermal transport experiments. They identified two separate transition temperatures, T* and $T_{JT}$. The position of T* was determined by a deviation from linear behaviour in a ln( /T) versus T plot, typically used to determine the activation energy for phonon assisted hopping. $T_{JT}$ is associated with a jump in Weiss constant and a large change in resistivity. Thus the transition was interpreted as follows: below T*, a long range (cooperatively) orbitally ordered phase exists. Between T* and $T_{JT}$ the volume fraction of an orbitally disordered phase with short range fluctuating Jahn Teller distortions starts to grow with temperature. Ultimately, the orbitally disordered phase is present above $T_{JT}$.

The situation for the smaller rare earth manganites and for the doped compounds is less clear and less studied. For the undoped compounds, the only structural studies are on $PrMnO_3$ (Sanchez *et al* 2002) and $NdMnO_3$ (Maris *et al*, 2004). The observed latent heat in DTA for $PrMnO_3$ (Sanchez *et al* 2002) and phase coexistence in $NdMnO_3$ as observed in diffraction experiments by Maris *et al*. provide evidence for a first-order phase transition. In contrast to these results, Zhou and Goodenough (2003) observed a crossover from a first-order to a second-order phase transition when they decreased the rare earth radius in transport measurements on single crystals of $LaMnO_3$, $PrMnO_3$ and $NdMnO_3$. A comparable effect was found by Bhattacharya *et al* (2004) who progressively substituted La by Nd at the A-site. The crossover from a first-order to second-order phase transition is reflected in a gradual change from a jump in resistivity associated with the orbital order-disorder phase transition in $LaMnO_3$ to a more continuous decrease in temperature dependent resistivity for smaller rare earths, combined with a decrease in latent heat for large Nd doping of $LaMnO_3$ using DSC experiments.

Divalent A-site doping replaces Jahn-Teller active $Mn^{3+}$ ions by Jahn-Teller inactive $Mn^{4+}$ ions and therefore disrupts the cooperative long-range orbital order mediated by the corner-shared oxygens of the octahedra. Chatterji *et al* have performed two studies on Calcium (2006) and Barium (2004) doped $LaMnO_3$ where it was shown that the magnitude of the discontinuous step in molar volume decreases and that the first-order phase transition gradually softens to become a (second-order-like) continuous phase transition.

Depending on the type of experiment, the exact nature of the phase transition is interpreted differently. The characterisation of the orbital order-disorder phase transition is a central issue in this paper. We report the effect of low Ca doping on the orbital ordering transition in $Nd_{1-x}Ca_xMnO_3$ based on high temperature X-ray powder diffraction measurements. A large temperature range of phase coexistence (>150K) associated with a first-order transition has been observed for low doping (x=0, 0.02). The phase fractions evolve continuously from the low temperature phase to the high temperature phase ressembling a martensitic transformation. At higher doping, the first-order nature of the transition and the phase coexistence are suppressed.
We show that non-thermodynamic probes such as response functions and transport experiments are difficult to interpret. In essence, for low doping transitions remain first-order in nature, evidenced by discontinuous temperature dependence of the lattice constants. However, the transition appears continuous because of a gradual evolution of the volume fractions of the phases over a broad temperature interval.

## Experimental

Polycrystalline samples of $Nd_{1-x}Ca_xMnO_3$ (x=0, 0.02, 0.05, 0.1) were prepared by conventional solid-state reaction by mixing stoichiometric amounts of $Nd_2O_3$, $CaCO_3$ and $MnO_2$. Phase purity was confirmed by powder X-ray diffraction (PXRD). Oxygen stoichiometry was checked by the redox titration method (Yang *et al* 2005). Variable temperature PXRD data were collected using a Huber diffractometer with Mo-$K_{\alpha 1}$ radiation. Several synchrotron PXRD datasets of $Nd_{1-x}Ca_xMnO_3$ (x=0.02 and 0.05) were collected on the BW5 diffractometer (Bouchard *et al* 1998) at Hasylab in Hamburg, Germany, using an X-ray wavelength of 0.12420 Å.. Data were collected between 1.5 ≤ 2θ ≤ 6.5 and binned with a step size 0.0025° for full structure refinement at 300K, 600K and 1250K. The powder samples were placed in a quartz capillary and heated by means of a thermal-image furnace. Temperature was measured by a thermocouple. The evolution of specific peaks with temperature was followed by scanning between 3.65 ≤ 2θ ≤ 3.8 in 20K steps. The GSAS suite of programs (Larson and Von Dreele 2000) was used for Rietveld fitting of the PXRD data.

## Results

*(a) Crystal structure at room temperature*
We have investigated the structure of stoichiometric $NdMnO_3$ and $Nd_{1-x}Ca_xMnO_3$ in the low doping range for x= 0.02, 0.05 and 0.10 by PXRD. All the samples were found to crystallize in an orthorhombic structure without extra diffraction peaks indicative of impurities. The room temperature crystal structure of all four samples is orthorhombic (space group *Pbnm*). Fig. 1(a) shows the evolution of the lattice parameters as a function of doping at room temperature from lab XRD data. Error bars are smaller than the symbols. All the samples have a ratio of unit cell parameters $c/\sqrt{2} < a < b$ corresponding to the so-called O'-orthorhombic structure (Goodenough *et al* 1961). The lattice parameter $a$ is hardly affected by the doping whereas the lattice parameter $b$ decreases strongly. The lattice parameter $c$ increases with increasing doping.

The JT distortion parameter $\Delta$ is defined as the variance of the Mn-O distances $d_n$ ($n$=1,6): $\Delta = (1/6) \sum_{n=1,6}[ |(d_n - <d>)|/<d>]^2$. The distortion modes characterizing the JT effect are defined as: $Q_2=2(l-s)/\sqrt{2}$ and $Q_3= 2(2m-l-s)/\sqrt{6}$ with $l$ and $s$ the long and short Mn-O distances in the *ab* plane and $m$ the medium out-of-plane Mn-O bond length (Kanamori 1960). Using these definitions, we plot in Fig. 1(b) the JT parameter $\Delta$ and the two octahedral distortion modes $Q_2$ and $Q_3$ as a function of Ca doping. In $Nd_{1-x}Ca_xMnO_3$, the JT-active $Mn^{3+}$ ions are replaced by non-JT active $Mn^{4+}$ ions upon Ca doping. Because of the corner-sharing nature of the perovskite structure, the JT parameter $\Delta$ decreases with increasing $Mn^{4+}$ content; it is reduced by one third for 10 % doping but not suppressed. The decrease of the JT distortion by doping is also reflected in the modes $Q_2$ and $Q_3$ that are also reduced by a factor of 2 at $x$=0.1.

(b) *Investigation of JT transition in $Nd_{1-x}Ca_xMnO_3$ at high temperature*

Figure 2 shows the thermodiffractograms of $Nd_{1-x}Ca_xMnO_3$ for $x$=0.02 (a) and $x$=0.05 (b) for a range of temperatures across $T_{JT}$, measured using lab PXRD. Figure 2 (a) illustrates the evolution of 220 peak of the low temperature (LT or O') and high temperature (HT or O) phases of the 2% calcium doped Nd manganite. The transition is discontinuous and we note that peaks corresponding to the high and low temperature phases coexist in a temperature range of at least 150 K. To illustrate how the Jahn-Teller transition evolves with doping, we show in fig. 2(b) a thermodiffractogram of $Nd_{1-x}Ca_xMnO_3$ for $x$=0.05. Here we choose to follow the 220 and 004 peaks close to the Jahn-Teller transition. In this case, the transition is more continuous and we do not observe phase coexistence. Rietveld refinement of the complete diffraction profiles yielded the lattice parameters as a function of temperature plotted for $x$=0.02 and $x$=0.05 in figure 3. Error bars are smaller than the symbols. We can distinguish two types of behavior depending on the doping level: for the low doping ($x$=0.02, panel a) we note that the lattice parameter $b$ decreases continuously with temperature and then drops abruptly at $T_{JT}$ whereas the lattice parameter $c/\sqrt{2}$ increases continuously and expands abruptly at $T_{JT}$. The lattice parameter $a$ is almost temperature independent. It is necessary to refine the diffraction profiles for temperatures from 620 to 770C using two phases. The volume fractions of the O' (low temperature) and O (high temperature) phases as a function of temperature are plotted in the inset. The phase fractions were calculated from the Rietveld refinement via the relation $f_p=(S_pV_p)/(\sum_{i=1}^{N} S_iV_i)$, where $p$ is the value of $i$ for a particular phase among the $N$ phases present, $S_i$ and $V_i$ are the refined scale factor and the volume of the unit cell for the $i^{th}$ phase, respectively. There is a continuous change in the phase fraction from the low temperature phase to the high temperature phase over more than 150K. For higher doping ($x$=0.05 and 0.1, panel b), the diffraction data can be modeled adequately by a single phase. The change in the lattice parameters is smooth, continuous and shows no abrupt change at the phase transition temperature.

## Discussion

The classification scheme of phase transitions makes a division into two broad categories based on the behavior of the temperature derivatives of the free energy. First-order phase transitions involve a discontinuous change in the intrinsic thermodynamic variables, such as molar volume, and are associated with latent heat. The second class of phase transitions are the 'continuous phase transitions', also called second-order phase transitions. These transitions do not involve latent heat but exhibit critical phenomena such as with order-disorder transitions. Phase coexistence arises only from first-order transitions. Electronic phase coexistence can thus be associated with metal-insulator or orbital ordering transitions. Phase coexistence implies that superheating/cooling effects are not dominant, and that the kinetics of the transition allows the formation of a state in thermodynamic equilibrium by adjusting the volume fraction continuously.

Chatterji *et al* (2006) use the unit cell volume as an order parameter of the phase transition in $La_{1-x}Ca_xMnO_3$. They observed a drop of the volume at $T_{JT}$. We prefer to use the lattice parameter *b* to locate the transition temperature: the long bonds in the JT-distorted phase are more oriented along *b* than along *a*, giving rise to an expansion of the unit cell along b below $T_{JT}$. It can thus be used as an order parameter to probe where $T_{JT}$ lies. Fig. 4 (a) shows the evolution of the lattice parameter *b* as a function of temperature for $Nd_{1-x}Ca_xMnO_3$ for x=0, 0.02, 0.05, 0.1. Error bars are smaller than the symbols. The discontinuity in *b* for x=0 and 0.02 is a signature of the first-order nature of the transition. This transition also presents a phase coexistence regime, clearly observed in the diffraction data. The diffraction peaks of the high temperature and low temperature phases are simultaneously present. The temperature $T_{JT}$ is defined as the middle of the phase coexistence regime. For x=0.05 and 0.1 the continuous development of *b* indicates a second-order transition. The transition temperature is defined as the onset of the deviation from the high temperature trend in *b*. All the transition temperatures as well as the limits of phase coexistence regime are displayed in figure 4 (b) as a function of doping.

The orbital ordering transition for $Nd_{1-x}Ca_xMnO_3$ in the low doping region (x=0, 0.02) is first order, as reflected in the jump in lattice parameters, and is similar to that observed in Ca doped $LaMnO_3$. The cross-over from first to second order with increase of doping has also been reported for Ca and Ba doped $LaMnO_3$ (Chatterji *et al* 2004, 2006). Maitra *et al* (2004) proposed that anharmonic coupling exists between the JT distortion (first order parameter) and volume strain (second order parameter). A reduction of this coupling would then lead to the observed cross-over behavior. However, this is a phenomenological theory and does not explain how Ca doping could reduce the anharmonic coupling parameter.

A striking feature of the Nd manganite series is that the regime of phase coexistence is particularly large: more than 200K for the undoped and more than 150K for the 2% doped sample. In comparison, for the La series this regime covers an interval of 10-30K (Chatterji *et al* 2006)). In the phase coexistence regime there is a continuous change of the phase fractions with temperature. This indicates thermodynamic equilibrium.

This transition presents some resemblance to martensitic transformations. Although some displacive phase transitions may be thermodynamically second or higher order, the martensitic transformation involves large distortions of the unit cell and is first order. The martensitic transformation is diffusionless, the change in crystal structure being achieved by a deformation of the parent phase and large changes in the lattice parameters. The shape deformation accompanying the transformation can be reversed by transforming back to the parent phase. Despite its first-order nature, the martensitic transition requires heating or cooling over a significant temperature interval to reach completion and is characterized by start and finish temperatures, $T_S$ and $T_F$. The nucleation and growth of domains of the transformed phase in a matrix of the untransformed phase can occur very quickly. However, when there is a mismatch of the lattice parameters at the interfaces of the transformed domains, long-range, anisotropic strain fields develop and hinder continued transformation in the rest of the sample. Martensites are recognizable by morphology as the transition proceeds preferably via strain-free interface planes. Although this type of transition is named after the steel alloy martensite, it has also been reported at the first-order charge-ordering transition in manganites at much larger dopings (Podzorov *et al* 2001, Kolb *et al* 2004, Lu *et al* 2006). It may also apply to the current systems, as in both cases the transition involves the onset of long-range orbital ordering; this happens together with charge ordering in the higher doped systems.

The analogy with martensites may indicate why sample to sample variations result in different observed behaviour. Most of the transport measurements in the literature have been performed on single crystal samples, whereas most of the diffraction experiments have been performed on powders. In powders there is a larger distribution of grain sizes and shapes (and possibly compositions) which gives rise to different parts of the sample undergoing the transition at different temperatures. That is, the energy barrier to the transition varies according to the grain morphology. This results in the phase coexistence region of the bulk sample being larger, but does not change the nature of the transition. A single crystal sample is much more "uniform". For instance in pure $LaMnO_3$, Prado *et al* (1999) have reported 100K of phase coexistence regime in a powder/polycrystalline sample whereas 10K is reported for crushed single crystal (Rodriguez-Carvajal *et al* 1998, Chatterji *et al* 2006), which is a polycrystalline material with a homogeneous composition. We note that a marked dependence of physical properties on grain size was observed in $LaMnO_3$ by Das *et al* 2006. Futhermore, shape memory effects that have been studied in thermoelastic martensitic metals were also observed in small rare-earth manganites ($RMnO_3$, R=Nd to Dy) by Kasper and Troyanchuk (1996).

If the orbital ordering transition in the current system for doping less than 2% is of martensitic nature, this may also help to explain the crossover to second-order character at higher dopings. The change in lattice parameters at the transition becomes much smaller with doping (Fig. 3) and thus strain will be reduced in the higher doped samples. Beyond a critical doping concentration between 2% and 5%, the structure may be able to accommodate the JT distortion in gradual fashion as it sets in, without the need to adjust for strain with the martensitic mechanism. However, further studies would be needed to demonstrate the presence of a martensitic transition in the current systems and conclusions cannot be made from diffraction data alone.

We note that phase coexistence involving orbital ordering in transition metal oxides can also have a different origin, as observed in $SmVO_3$ (Sage *et al* 2006). There, the phase coexistence of orbital orderings of different symmetry sets in at low temperature (115K) where both phases initially have the same molar volume. The phase coexistence is the result of magnetic exchange striction, which provides the coupling to the lattice.

In samples exhibiting electronic phase coexistence it is easy to misinterpret thermodynamic and transport measurements. They measure the average properties of the bulk sample and thus do not probe the properties of the individual phases. This explains why the orbital order-disorder transition in manganites has been assigned as either first or second order, depending on the type of experiment. For instance, in pure $PrMnO_3$ and $NdMnO_3$ the transition is seen $1^{st}$ order by diffraction (Maris *et al* 2004) but $2^{nd}$ order by transport (Zhou and Goodenough 2003) and latent heat (Battacharya *et al* 2004). The picture of a phase-coexistence region with continuously varying phase fractions reconciles these two points of view. Furthermore, our measurements show that the two consecutive phase transitions $T_{JT}$ and $T^*$ found by transport and thermoelectric power measurements (Zhou and Goodenough 2003) should be interpreted as the two temperatures defining the phase coexistence regime of the unique first-order orbital ordering transition. To conclude, non-thermodynamic probe measurements of the JT transition at low doping should be interpreted as first-order with phase coexistence. For a much higher doping the transition is second order. The calcium concentration $x_c$ of the cross-over between first and second-order is $0.075 > x_c > 0.1$ for the La serie and $0.02 > x_c > 0.05$ for the Nd serie. There is too few data to delineate the rare-earth dependance of the cross-over concentration.

## Conclusion

In summary, we show experimental evidence of the cross-over from a first to a second-order transition with doping for the onset of orbital ordering in low-doped $Nd_{1-x}Ca_xMnO_3$. A large temperature range of phase coexistence (>150K) associated with the first-order transition has been observed for low doping (x=0, 0.02) and has features characteristic of a martensitic transition. The phase fraction evolves continuously from the low temperature phase to the high temperature phase. At higher doping, the first-order nature of the transition and the phase coexistence are suppressed.

## Acknowledgements

We thank G.R Blake for helpful discussions. This work is supported by the European Community under Grant COMEPHS No. NMPT4-CT-2005-517039.

**Figure captions**:

Fig. 1. Room temperature doping variation of the (a) lattice parameters of $Nd_{1-x}Ca_xMnO_3$ (b) JT distortion parameter $\Delta$ and the two octahedral distortion modes $Q_2$ and $Q_3$ (inset) defined in the text.

Fig. 2. X-ray thermodiffractogram of $Nd_{1-x}Ca_xMnO_3$ for x=0.02 (a) and x=0.05 (b) showing the diffraction peaks of the low temperature (LT or O') and high temperature (HT or O) phases and their temperature dependence.

Fig. 3. (a) Temperature variation of the lattice parameters of $Nd_{1-x}Ca_xMnO_3$ for x=0.02. The filled and open symbols correspond to the low temperature O' and high temperature O phases, respectively. The hatched area represents the range of temperature of phase coexistence. The inset shows the fraction of the O' and O phases as a function of temperature. (b) Temperature variation of the lattice parameters of $Nd_{1-x}Ca_xMnO_3$ for x=0.05. The dashed lines show the extrapolation of high temperature dependance of the lattice parameters.

Fig. 4. (a) Temperature variation of the lattice parameter b of $Nd_{1-x}Ca_xMnO_3$ for x=0, 0.02, 0.05, 0.1. The hatched areas represent the temperature of phase coexistence. The arrows mark orbital ordering transition temperature $T_{JT}$. (b) Doping dependence of the orbital ordering transition temperature $T_{JT}$. The upper and lower limits of the phase coexistence regime are indicated by bars. The hatched area represents phase coexistence regime.

**Figure 1**:
(a)

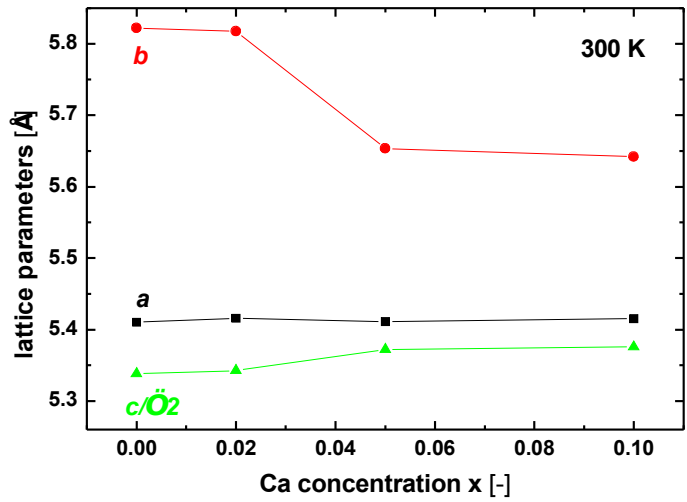

(b)

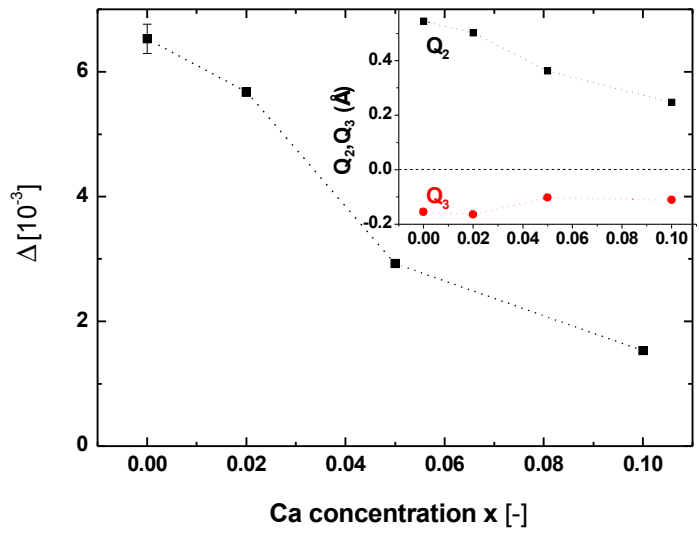

**Figure 2**:
(a)

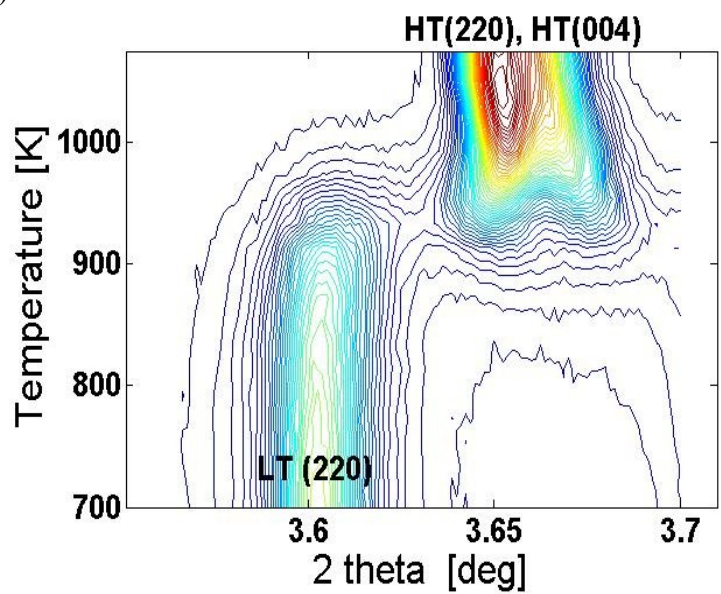

(b)

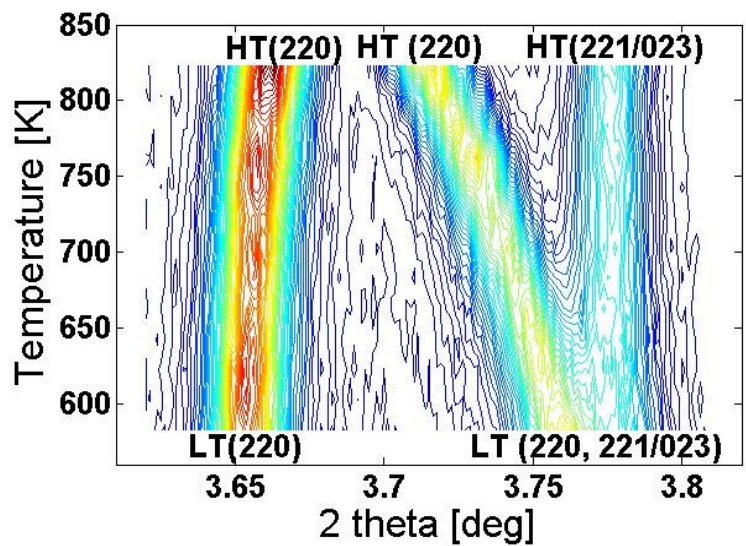

**Figure 3**: (a)

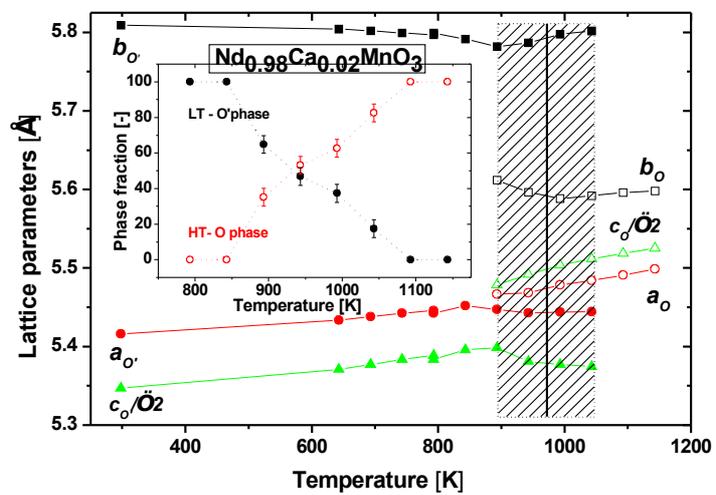

(b)

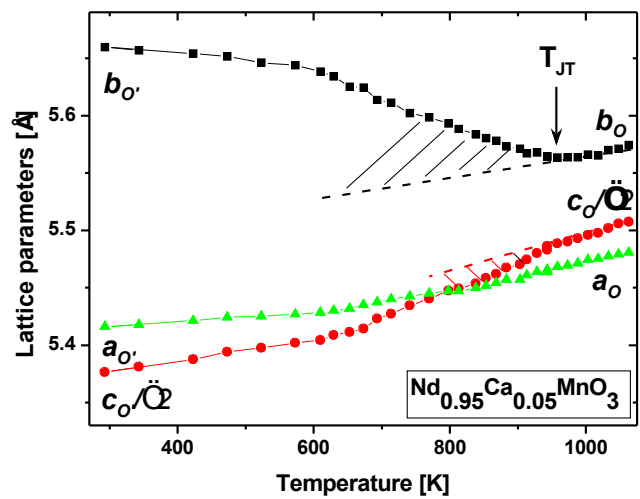

**Figure 4**: (a)

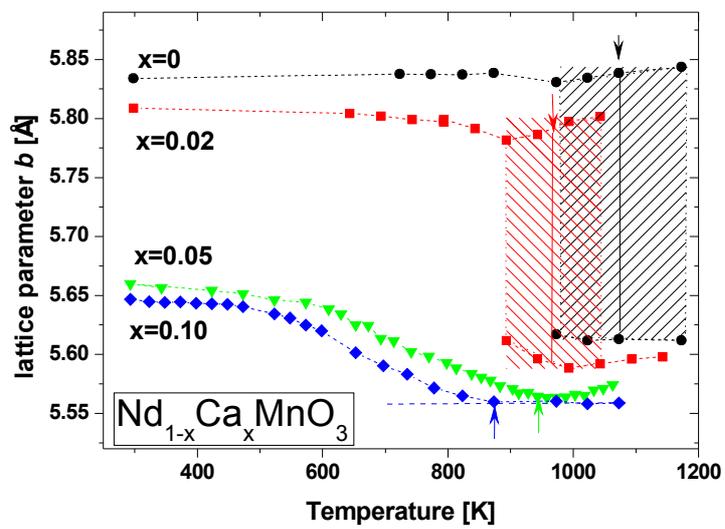

(b)

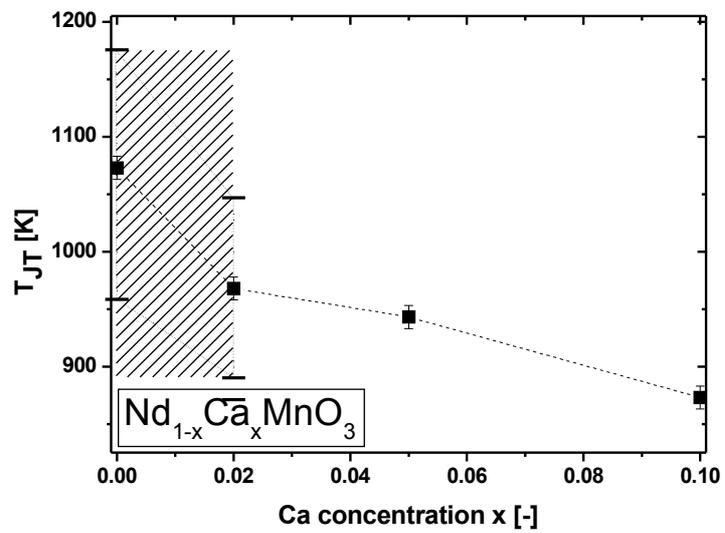

# References

Bhattacharya D, Devi P S and Maiti H S 2004 *Phys. Rev.* B **70** 184415

Bouchard R, Lippmann T, Neuefeind J, Neumann H-B, Poulsen H F, Rütt U, Schmidt T, Schneider J R, Süssenbach J and v. Zimmermann M 1998 *Journal of Synchrotron Radiation* **5** 90-101

Chatterji T, Fauth F, Ouladdiaf B, Mandal P, and Ghosh B 2003 *Phys. Rev.* B **68** 052406

Chatterji T, Ouladdiaf B, Mandal P, and Ghosh B 2004 *Solid State Commun.* **131** 75

Chatterji T, Riley D, Fauth F, Mandal P and Ghosh B 2006 *Phys. Rev.* B **73** 094444

Das N, Mondal P, Bhattacharya D 2006 *Phys. Rev.* B **74** 014410

Goodenough J B, Wold A, Arnott R J and Menyuk N 1961, *Phys. Rev.* **124** 373

Kanamori J 1960 *J. Appl. Phys.* **31** 14S

Kasper N V and Troyanchuk I O 1996 *J. Phys. Chem Solids* **57** 1601-1607

Kolb P W, Romero D B, Drew H D, Moritomo Y, Souchkov A B and Ogale S B 2004 *Phys. Rev.* B **70** 224415

Larson A C and Von Dreele R B 2000 Los Alamos National Laboratory Report LAUR **86**-748

Lu W J, Sun Y P, Zhao B C, Zhu X B and Song W H 2006 *Phys. Rev.* B **73** 214409

Maitra T, Thalmeier P and Chatterji T 2004 *Phys. Rev.* B **69** 132417

Maris G, Volotchaev V and Palstra T T M 2004 *New Journal of Physics* **6** 153

Mathur N and Littlewood P 2003 *Physics Today* january 2003 25

Podzorov V, Kim B G, Kiryukhin V, Gershenson M E and Cheong S-W 2001 *Phys. Rev.* B **64** R140406

Prado F, Sanchez R D, Caneiro A, Causa M T, and Tovar M 1999b *J. Solid State Chem.* **146** 418-427

Prado F, Zysler R, Morales L, Caneiro A, Tovar M, Causa M T 1999a *J. Magn. Magn. Mater.* **196-197** 481-483

Qiu X, Proffen Th, Mitchell J F, and Billinge S J L 2005 *Phys. Rev. Lett.* **94** 177203

Rodríguez-Carvajal J, Hennion M, Mousa F and Moudden A H 1998 *Phys. Rev. **B** * **57** R3189

Sage M H, Blake G R, Nieuwenhuys G J, and Palstra T T M 2006 *Phys. Rev. Lett.* **96** 036401

Sánchez D, Alonso J A and Martínez-Lope M J 2002 *J. Chem. Soc.*, *Dalton Trans*. 4422–4425

Sánchez M C, Subías G, García J, and Blasco J 2003 *Phys. Rev. Lett.* **90** 045503

Tokura Y and Nagaosa N 2000 *Science* **288** 462
Van Aken B B *et al* 2003 *Phys. Rev. Lett.* **90** 066403

Yang J, Song W H, Ma Y Q, Zhang R L and Sun Y P 2005 *J. Magn. Magn. Mater.* **285** 417

Zhou J-S and Goodenough J B 2003 *Phys. Rev.* B **68** 144406